\documentclass{aa}
\usepackage{graphics,epsfig,txfonts}

\usepackage{natbib}
\bibpunct{(}{)}{;}{a}{}{,}

\newcommand{\ergcm}[1]{$\times 10^{#1}$ erg cm$^{-2}$ s$^{-1}$}

\newcommand{\ergs}[1]{$\times 10^{#1}$ erg s$^{-1}$}

\newcommand{\hcm}[1]{$\times 10^{#1}$ cm$^{-2}$}

\newcommand{\nh}{N$_{\rm H}$}

\newcommand{\ct}{cts s$^{-1}$}

\newcommand{\Halp}{H${\alpha}$}
\newcommand{\ltsima}{$\buildrel < \over \sim$}
\newcommand{\lsim}{\lower.5ex\hbox{\ltsima}}
\newcommand{\gtsima}{$\buildrel > \over \sim$}
\newcommand{\gsim}{\lower.5ex\hbox{\gtsima}}
\newcommand{\xmm}{XMM-Newton}
\newcommand{\xmmp}{\hbox{XMMU\,J004814.1-731003}}
\newcommand{\xmms}{\hbox{J0048}}
\newcommand{\axj}{\hbox{AX\,J0048.2-7309}}
\newcommand{\hma}{\hbox{[MA93]\,215}}

\begin{document}
 
\title{XMM-Newton observations of the Small Magellanic Cloud:\\ 
       XMMU\,J004814.1-731003, a 25.55 s Be/X-ray binary pulsar
       \thanks{Based on observations with 
               XMM-Newton, an ESA Science Mission with instruments and contributions 
               directly funded by ESA Member states and the USA (NASA)}}
\author{F.~Haberl\inst{1} \and P.~Eger\inst{1} \and W.~Pietsch\inst{1} \and R.H.D.~Corbet\inst{2} \and M.~Sasaki \inst{1}}

\titlerunning{A 25.55 s Be/X-ray binary pulsar in the SMC}
\authorrunning{Haberl et al.}

\institute{Max-Planck-Institut f\"ur extraterrestrische Physik,
           Giessenbachstra{\ss}e, 85748 Garching, Germany\\
	   \email{fwh@mpe.mpg.de}
	   \and
	   University of Maryland, Baltimore County; X-ray Astrophysics Laboratory, Mail Code 662,
           NASA Goddard Space Flight Center, Greenbelt, MD 20771, USA
	   }
 
\date{Received 8 February 2008 / Accepted 6 March 2008}
 
\abstract{} 
         {To investigate candidates for Be/X-ray binaries in the 
          Small Magellanic Cloud (SMC), we observed a region around 
	  the emission nebula N19 with \xmm\ in October 2006.}
	 {We analysed the EPIC data of the detected point sources 
	  to derive their spectral and temporal characteristics.}
         {We detected X-ray pulsations with a period of 25.550(2) s
	  from the second-brightest source in the field, which we
	  designate \xmmp. The X-ray spectrum is well modelled by a
	  highly absorbed (\nh = 5\hcm{22}) powerlaw with photon 
	  index 1.33$\pm$0.27. The precise X-ray position allows us to
	  identify a Be star as the optical counterpart. \xmmp\ is 
	  located within the error circle of the transient ASCA source \axj, 
	  but its position is inconsistent with that of the
	  proposed optical counterpart of \axj\ (the emission line star \hma). 
	  It remains unclear if 
	  \xmmp\ is associated with \axj. \xmmp\ might be 
	  identical to an RXTE pulsar that was discovered with a period
	  of 25.5 s, but which is listed as 51 s pulsar in the 
	  recent literature. }
         {}

\keywords{galaxies: individual: Small Magellanic Cloud --
          galaxies: stellar content --
          stars: emission-line, Be -- 
          stars: neutron --
          X-rays: binaries}
 
\maketitle
 
\section{Introduction}

The region around the emission nebula N19 in the southwestern part of the 
SMC shows complex X-ray emission
from several supernova remnants and high mass X-ray binaries (HMXBs).
Several X-ray pulsars which are most likely associated with Be/X-ray binaries, the major
subclass of HMXBs, were detected with ASCA and RXTE, but could not be accurately
located and therefore, lack a clear optical identification.
Southeast of N19, ASCA detected a 9.1~s pulsar \citep{2000PASJ...52L..63U} where ROSAT
PSPC images show two close hard X-ray sources, both possibly contributing to the
ASCA source \citep{2000A&A...361..823F}. Both these objects are most likely
Be/X-ray binaries and it is not clear which is the 9.1 s pulsar
\citep[][ and references therein]{2005MNRAS.356..502C}. Another nearby 
ASCA source (\axj) was classified as candidate Be/X-ray binary by 
\citet{2003PASJ...55..161Y} from its X-ray spectral properties.
In the error circle of \axj\ the \Halp\ emission line star \hma\ 
\citep{1993A&AS..102..451M} is located, suggesting
an association with a Be star. However, this requires confirmation by an improved 
X-ray position. ROSAT did not detect any source at the ASCA position 
but a faint source was seen during an early \xmm\ observation in August 2002
\citep{2005MNRAS.362..879S}. RXTE discovered pulsations with periods of 
16.6 s and 25.5 s during an observation in September 2000, viewing the area
near the southwestern edge of the SMC \citep{2002ApJ...567L.129L}. The 
pulse period from the latter source was later
argued to be the harmonic of a 51 s spin period \citep{2005ApJS..161...96L}.
Both pulsars could never be better localized and optically identified.

In the course of our AO5 \xmm\ program to investigate candidates for HMXBs 
in the SMC, we observed the region around N19. The \xmm\ source detected 
by \citet{2005MNRAS.362..879S} was seen again and we analysed the EPIC data
to investigate the nature of this object, designated \xmmp.
Here we report on results from a temporal and spectral analysis of the X-ray data
of \xmmp\ and identify as optical counterpart a Be star.
This adds \xmmp\ to the numerous Be/X-ray binary pulsars known in the SMC
\citep[for recent reviews see][]{2004A&A...414..667H,2005MNRAS.356..502C}.

\section{Data analysis and results}

We observed the field around the emission nebula N19 with \xmm\ \citep{2001A&A...365L...1J}
on 2006 October 5. The EPIC-MOS \citep{2001A&A...365L..27T} and EPIC-PN 
\citep{2001A&A...365L..18S} cameras were operated in imaging mode 
(see Table~\ref{tab-obs}). 
For the X-ray analysis we used the \xmm\ Science Analysis System (SAS) version 7.1.0 
supported by tools from the FTOOL package together with XSPEC version 11.3.2p 
for spectral modelling.  

A source, not seen by ROSAT was detected near the center of the field of view
and inside the error circle of the ASCA source \axj\ \citep{2003PASJ...55..161Y}.
After astrometric boresight correction \citep[see also][]{2008arXiv0801.4679H} 
we determined the position of the source to
R.A. = 00 48 14.10 and Dec. = $-$73 10 04.0 (J2000.0) using the SAS standard maximum 
likelihood technique for source detection and assign the name \xmmp\ (hereafter \xmms). The 
1$\sigma$ position error is 1.2\arcsec, dominated by the remaining 
systematic uncertainty of 1.1\arcsec. 
Due to background flaring activity during part of the observation
we applied a 
background screening before our source detection analysis, but used the full 
exposure times for the extraction of light curves and spectra of individual 
sources. The resulting net exposures are listed in Table~\ref{tab-obs}.

\label{sect-obs}
\begin{table}
\caption[]{Details of the \xmm\ EPIC observation.}
\begin{tabular}{cccc}
\hline\hline\noalign{\smallskip}
\multicolumn{1}{c}{Observation} &
\multicolumn{2}{c}{Pointing direction} &
\multicolumn{1}{c}{Sat.} \\

\multicolumn{1}{c}{ID} &
\multicolumn{1}{c}{R.A.} &
\multicolumn{1}{c}{Dec.} &
\multicolumn{1}{c}{Rev.} \\

\multicolumn{1}{c}{} &
\multicolumn{2}{c}{(J2000.0)} &
\multicolumn{1}{c}{} \\
\noalign{\smallskip}\hline\noalign{\smallskip}
 0404680101 & 00 47 36.0 & -73 08 24.0 & 1249  \\

\noalign{\smallskip}\hline\noalign{\smallskip}
\multicolumn{1}{c}{EPIC$^{(1)}$} &
\multicolumn{1}{c}{Start time} &
\multicolumn{1}{c}{End time} &
\multicolumn{1}{c}{Net$^{(2)}$} \\

\multicolumn{1}{c}{instrument} &
\multicolumn{2}{c}{2006-10-05 (UT)} &
\multicolumn{1}{c}{exposure} \\

\multicolumn{1}{c}{configuration} &
\multicolumn{1}{c}{} &
\multicolumn{1}{c}{} &
\multicolumn{1}{c}{ks} \\
\noalign{\smallskip}\hline\noalign{\smallskip}
 PN FF thin   & 00:44:47 & 06:51:09 & 19.04 / 6.84 \\
 M1 FF medium & 00:22:05 & 06:50:49 & 22.76 / 7.66 \\
 M2 FF medium & 08:22:05 & 06:50:54 & 22.77 / 7.67 \\
\noalign{\smallskip}\hline\noalign{\smallskip}
\end{tabular}

$^{(1)}$ FF: full frame CCD readout mode with 73 ms frame time for PN and 2.6 s for MOS; 
     thin and medium optical blocking filters. $^{(2)}$
     Left: Full exposure times as used for
     spectral and temporal analysis. Right: Exposure after removing intervals of high
     background (used for source detection analysis, see text).
\label{tab-obs}
\end{table}

\begin{figure}
  \resizebox{0.98\hsize}{!}{\includegraphics[clip=]{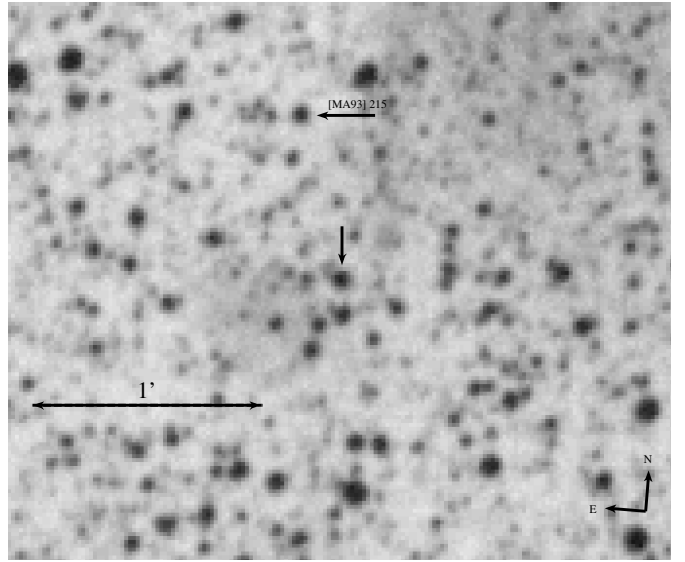}}
  \caption{Finding chart based on the DSS2 red image of the region around \xmmp.
           The optical counterpart is marked with a vertical arrow.
	   The error circle of the
	   X-ray position is smaller than the size of the star on the image.
	   The \Halp\ emission line star [MA93] 215 is also labelled.}
  \label{xmmp-fc}
\end{figure}

The precise X-ray position inferred from the EPIC data allowed us to identify the optical 
counterpart of \xmms. The source position is incompatible with the emission line star
\hma\ (46\arcsec\ away from the X-ray position) 
which was suggested as optical counterpart for \axj, but another star with optical
properties of a B star is found at the \xmm\ X-ray position (Fig.~\ref{xmmp-fc}).
In Table~\ref{tab-ids} optical brightness and colours taken from the
UBVR CCD Survey of the Magellanic Clouds \citep{2002ApJS..141...81M}, 
the Magellanic Clouds Photometric Survey \citep[MCPS,][]{2002AJ....123..855Z} and 
the OGLE BVI photometry catalogue \citep{1998AcA....48..147U} are given.

\begin{figure}
  \resizebox{0.98\hsize}{!}{\includegraphics[angle=-90,clip=]{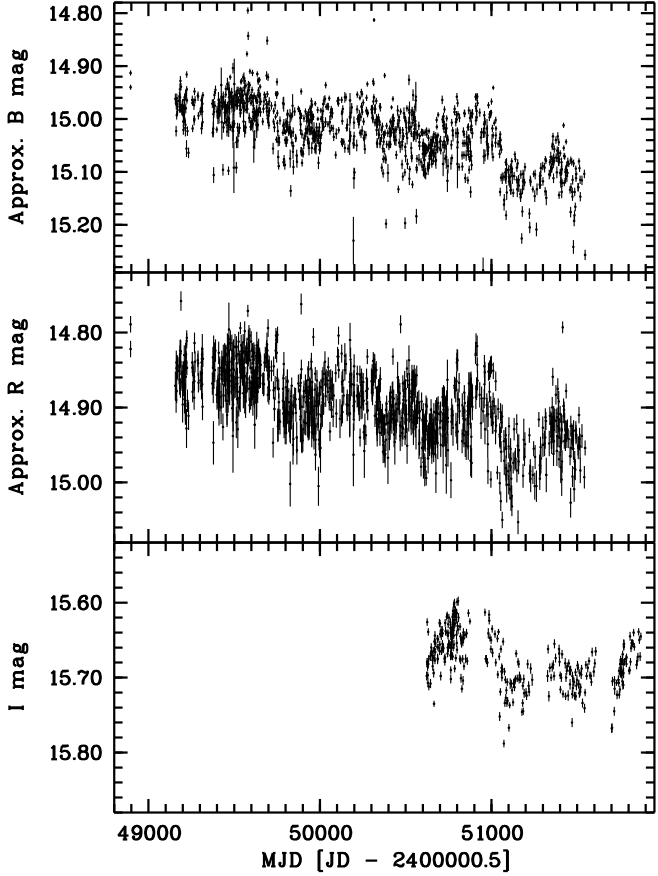}}
  \caption{MACHO B- and R- and OGLE I-band light curves of the optical counterpart of 
   \xmmp. For the MACHO data only data points with errors less than 0.05 mag are drawn.}
  \label{xmmp-opt}
\end{figure}

\begin{figure}
  \resizebox{0.98\hsize}{!}{\includegraphics[clip=]{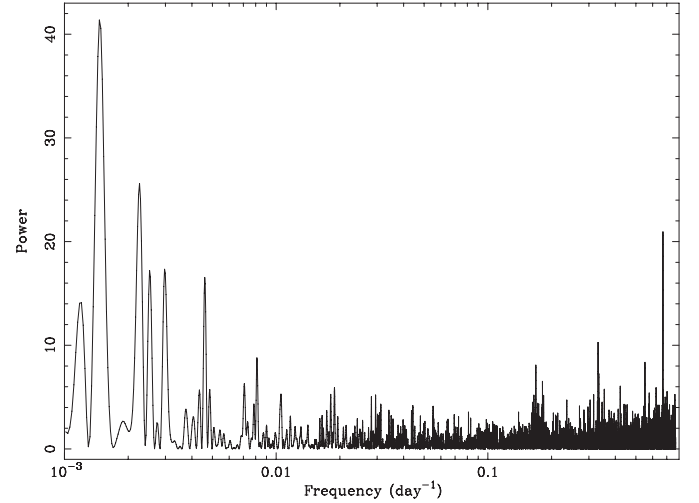}}
  \caption{Power spectrum of the MACHO R-band data of the optical counterpart 
           (star 212.15849.52).}
  \label{xmmp-redfft}
\end{figure}

To investigate longterm brightness variations of the optical counterpart 
we retrieved light curves in the I-band from the OGLE photometry database 
\citep[star 171264;][]{2005AcA....55...43S,1997AcA....47..319U}
and the B- and R-band from the MACHO survey (star 212.15849.52).
The light curves in all bands show a gradual fading over the total observing period 
with additional variations by $\sim$0.2 magnitudes superimposed (Fig.~\ref{xmmp-opt}). 
We applied an FFT analysis to the MACHO R- and B-band data with a magnitude 
error $< 0.05$~mag \citep{1976Ap&SS..39..447L,1982ApJ...263..835S} to 
determine whether these changes are periodic. 
We investigated the period range 
between 1 day and 1000 days which revealed peaks at short and long periods (Fig.~\ref{xmmp-redfft}).
Peaks at 1.49 days, 1.81 days, 3.02 days and 5.93 days indicate similar periods as seen from
other SMC Be/X-ray binaries \citep{2004AJ....127.3388S}, but are probably too 
short for orbital periods.
Broad peaks at long periods of a few hundred 
days ($\sim$220 days, $\sim$340 days and $\sim$400 days, $\sim$440 days and $\sim$ 680 days)
indicate variations which are not strictly periodic.

\begin{table*}
\caption[]{Optical identification.}
\begin{center}
\begin{tabular}{llcccccc}
\hline\hline\noalign{\smallskip}
\multicolumn{1}{l}{Source} &
\multicolumn{1}{l}{Catalogue} &
\multicolumn{1}{c}{R.A. and Dec. (J2000.0)} &
\multicolumn{1}{c}{Vmag} &
\multicolumn{1}{c}{B$-$V} &
\multicolumn{1}{c}{U$-$B} &
\multicolumn{1}{c}{V$-$R} &
\multicolumn{1}{c}{V$-$I} \\

\noalign{\smallskip}\hline\noalign{\smallskip}
\xmmp & UBVR & 00 48 14.10 --73 10 04.0 & 15.25 & $+$0.13 & $-$0.64 & $+$0.12 & --      \\
      & MCPS & 00 48 14.18 --73 10 03.9 & 15.30 & $+$0.26 & $-$0.50 & --      & $-$0.21 \\
      & OGLE & 00 48 14.13 --73 10 03.5 & 15.71 & $+$0.01 & --	    & --      & $+$0.09 \\
\noalign{\smallskip}\hline
\end{tabular}
\end{center}
\label{tab-ids}
\end{table*}

\begin{figure}
  \resizebox{0.98\hsize}{!}{\includegraphics[clip=]{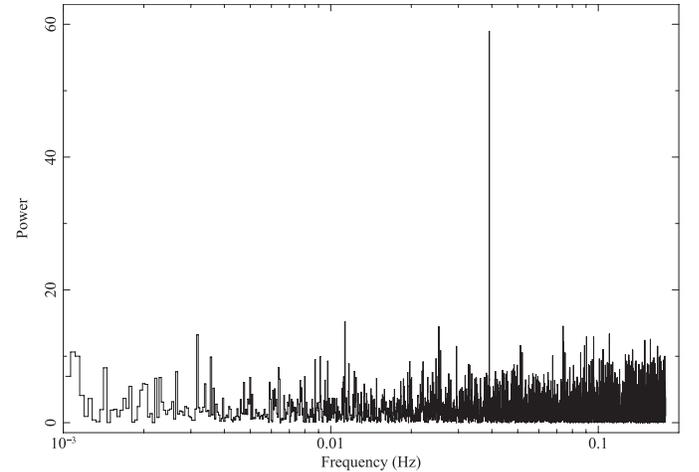}}
  \caption{Power spectrum of \xmmp\ produced from the 1.0-10.0 keV combined EPIC data.}
  \label{xmmp-pspec}
\end{figure}

The broad band X-ray light curve of \xmms\ indicates some intensity variations of
the flux on time scales of 1$-$2 hours. But due to the low count rate
(on average $\sim$0.03 \ct\ in EPIC-PN) the statistics is too low for 
definite conclusions.
We searched for X-ray pulsations in the EPIC data of \xmms, after correcting
the photon arrival times to the solar system barycenter. The Fourier power spectra
obtained from the EPIC-PN data in different energy bands revealed a peak at 25.5 s
which is most significant when selecting energies above 1 keV. For the band 1.0$-$10.0 keV a power 
of $\sim$40 is obtained. To verify the period, also the MOS data were investigated. 
Peaks at the same frequency were present for the individual MOS light curves, but not
significant on their own. Finally, combining the events from the two MOS cameras, also resulted in
the highest peak at the same frequency. Combining all EPIC data yields a maximum power of
$\sim$60 (Fig.~\ref{xmmp-pspec}).
Although no significant power around 51 s is seen in the power spectrum, we investigated this
period range using the Rayleigh $Z^2$ method \citep{1983A&A...128..245B} 
with fundamental and one harmonic frequency involved.
The resulting maximum $Z^2$ power is 69.5 with $Z^2_1$=2.9 for the fundamental (at 51.1 s) and $Z^2_2$=66.6 
for the first harmonic (at 25.55 s). We conclude that there is no significant modulation
with a period of 51 s.
To derive an accurate value and error for the detected pulse period we 
used the Bayesian method \citep{1996ApJ...473.1059G} as described in 
\citet{2000ApJ...540L..25Z}. The pulse period is determined to 25.550$\pm$0.002 s 
(1$\sigma$ error).

\begin{figure}
  \resizebox{0.98\hsize}{!}{\includegraphics[clip=]{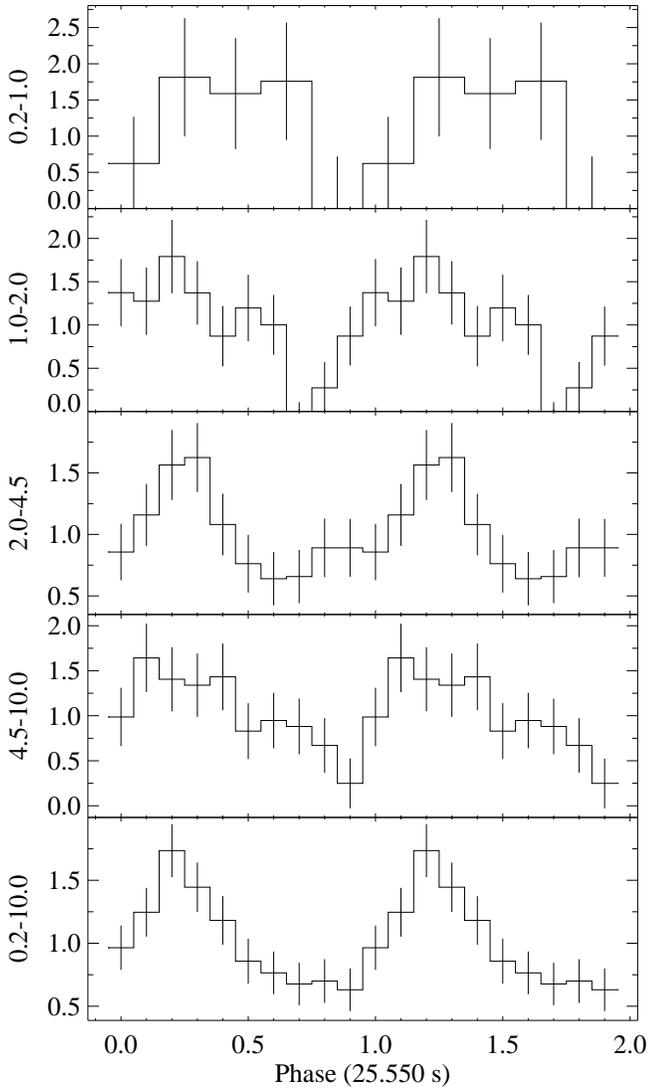}}
  \caption{Folded EPIC-PN light curves of \xmmp\ in the standard EPIC energy bands (the lower energy bands
           0.2$-$0.5 keV and 0.5$-$1.0 keV were combined and rebinned by a factor of two due to the 
	   low statistics). The panels
           show the pulse profiles for the different energies specified in keV. The intensity profiles 
           are background subtracted and normalized to the average count rate 
	   (in cts s$^{-1}$: 0.00082, 0.0073, 0.0135, 0.0093, 0.0310 from top to bottom).}
  \label{xmmp-pulse}
\end{figure}

We folded the X-ray light curves on the best period in the standard EPIC energy bands 
(0.2$-$0.5 keV, 0.5$-$1.0 keV, 1.0$-$2.0 keV, 2.0$-$4.5 keV, 4.5$-$10.0 keV and the
broad band 0.2$-$10.0 keV) and show the resulting pulse profiles in Fig.~\ref{xmmp-pulse}.
The pulse profile of \xmms\ is dominated by a broad main pulse which 
varies in shape with energy. At energies below 1 keV the high absorption (see below) 
strongly reduces the count rate and a modulation is only marginally seen.
The pulsed fraction (derived from modelling the pulse profile with two sine waves)
in the total energy band is (60$\pm$20)\%. There is some indication
for a higher pulsed fraction at energies below 1 keV and above 4.5 keV, but the number 
of counts in the individual bands is insufficient for a statistically justified statement.

\begin{figure}
  \resizebox{0.98\hsize}{!}{\includegraphics[angle=-90,clip=]{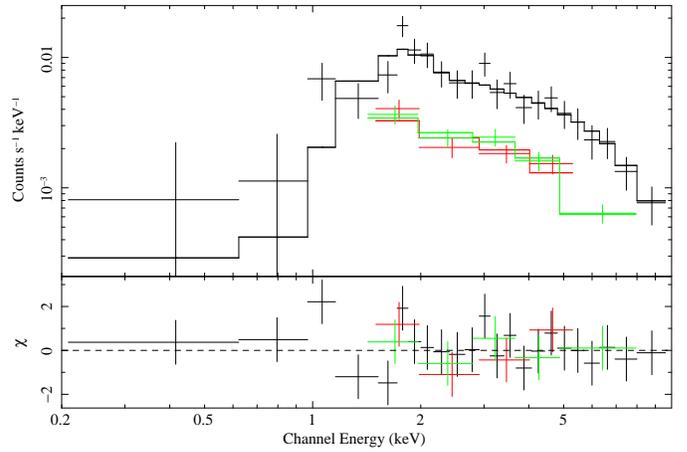}}
  \caption{EPIC spectra of \xmmp. EPIC-PN is shown in black and 
  EPIC-MOS in red (M1) and green (M2) (both grey in black and white representation). The histograms show
  the best-fit absorbed powerlaw model.}
  \label{xmmp-spec}
\end{figure}

For the analysis of the X-ray spectra we extracted pulse-phase averaged EPIC spectra for 
PN (single + double pixel events, PATTERN 0$-$4) and MOS (PATTERN 0$-$12) 
disregarding bad CCD pixels and columns (FLAG 0). 
The three EPIC spectra were simultaneously fit with an absorbed powerlaw model
allowing for a constant normalization factor between the spectra. 
We used two absorption components, accounting for the Galactic foreground absorption 
\citep[with a fixed hydrogen column density of 6\hcm{20} and elemental abundances from][]{2000ApJ...542..914W}
and the SMC absorption 
\citep[with column density as free parameter in the fit and with metal abundances reduced to 
       0.2 as typical for the SMC;][]{1992ApJ...384..508R}.
The best-fit (reduced $\chi^2$ = 0.82 for 27 degrees of freedom) 
powerlaw model yields a high absorption with \nh\ = (5.2$\pm$2.0) \hcm{22},
a photon index of 1.33$\pm$0.27, an observed flux of 3.5\ergcm{-13} and a source
luminosity of 2.1\ergs{35} 
\citep[0.2$-$10.0 keV, assumed distance to the SMC of 60 kpc][]{2005MNRAS.357..304H}.
Flux and luminosity refer to the values derived from the EPIC-PN spectrum, MOS1 and MOS2 yield
somewhat lower values by 6\% and 3\%, respectively. Errors for spectral parameters denote 
90\% confidence limits. 
The EPIC spectra with the best-fit model are shown in Fig~\ref{xmmp-spec}.

\section{Discussion}

We report the discovery of pulsations in the X-ray flux of \xmms\ with a period of 25.550 s
and propose as optical counterpart a V$\sim$15.5 mag star. 
The X-ray spectrum of \xmms\ is well represented by a simple powerlaw, typical for
Be/X-ray binary pulsars. However, the high absorption and the relative faintness of the
source (the flux of 3.5\ergcm{-13} corresponds to a source luminosity of 2.1\ergs{35} 
at SMC distance) prevent a more detailed spectral analysis.

Following 
\citet{2005MNRAS.356..502C}, we estimate the spectral class of the optical counterpart from 
its B$-$V colour index to B3 or later, depending on the 
observed B$-$V index which is in the range 0.01 to 0.26 (Table~\ref{tab-ids}). 
This assumes an extinction correction to the SMC of E(B$-$V)=0.08 \citep{1991A&A...246..231S} 
and an additional correction of (B$-$V)=$-$0.13 to account for the presence of a
circumstellar disc \citep{2005MNRAS.356..502C}. However, in the Galaxy no Be/X-ray binaries 
with spectral type later than B2 are found \citep{1998A&A...338..505N} which may 
indicate that this additional correction for
\xmms\ is insufficient. This is supported by the large absorption seen in the X-ray spectrum of 
the pulsar. The significantly different values measured for B$-$V in the different 
photometric surveys also suggest, that the extinction varies
with time (also 
a B$-$R colour index derived from the approximate B and R 
light curves in Fig.~\ref{xmmp-opt} slowly increased
after the end of 1998). At least this variable extinction must originate from matter 
local to the binary system and is most likely due to changes in the disc of the Be star.

From the relation between spin and orbital period \citep[for a recent version of the `Corbet'
diagram of SMC pulsars see][]{2005AJ....130.2220S}, we expect an orbital period between 
$\sim$25 days and $\sim$150 days for \xmms. The optical light curves indicate various 
periodicities, which are either shorter or longer than this period range. 
Short periods between 3 and 11 days were also seen in other SMC Be/X-ray binaries 
and might be caused by a changing view of the Be disk region that is brightened by the 
neutron star \citet{2004AJ....127.3388S}.
The star shows optical brightness variations on long time scales of a few hundred days 
which most likely are also associated with the Be star phenomenon, similar to e.g.  
XTE\,J0103-728 \citep[=SXP6.85;][]{2008MNRAS.384..821M}, but with smaller amplitude.

The optical and X-ray properties identify \xmms\ as Be/X-ray binary pulsar in the SMC.
The \xmm\ source position is located within the error circle of \axj, but the presence of another 
Be star (\hma), which might be associated with \axj, leaves it unclear if the ASCA source is 
identical with \xmms. Only a retrospective detection of the pulse period in the ASCA 
data would allow an unambiguous identification.
The pulsar might be identical to the 25.55s pulsar seen by RXTE in 
September 2000 \citep{2002ApJ...567L.129L}. The RXTE observation was pointed
at R.A. = 00 50 44.64 and Dec. = $-$73 16 04.8 which is 12\arcmin\ away from \xmms.
Therefore \xmms\ was well within the RXTE PCA field of view of 1\degr\ FWHM.
However, in more recent literature \citep{2005ApJS..161...96L,2005MNRAS.356..502C,2008arXiv0802.2118G}
the RXTE pulsar is listed with a period of 51 s. 
Therefore, we can not exclude the possibility that the \xmm\ and RXTE sources 
are two pulsars with different period.

\begin{acknowledgements}
The XMM-Newton project is supported by the Bundesministerium f\"ur Wirtschaft und 
Technologie/Deutsches Zentrum f\"ur Luft- und Raumfahrt (BMWI/DLR, FKZ 50 OX 0001)
and the Max-Planck Society. 
The ``Second Epoch Survey'' of the southern sky was produced by the 
Anglo-Australian Observatory (AAO) using the UK Schmidt Telescope. 
Plates from this survey have been digitized and compressed by the ST ScI. 
Produced under Contract No. NAS 5-26555 with the National Aeronautics 
and Space Administration.
This paper utilizes public domain data obtained by the MACHO Project, jointly 
funded by the US Department of Energy through the University of California, 
Lawrence Livermore National Laboratory under contract No. W-7405-Eng-48, by 
the National Science Foundation through the Center for Particle Astrophysics 
of the University of California under cooperative agreement AST-8809616, and 
by the Mount Stromlo and Siding Spring Observatory, part of the Australian 
National University.
\end{acknowledgements}

\bibliographystyle{aa}
\bibliography{general,myrefereed,mcs,hmxb,ism,ins,cv}

\end{document}